\documentstyle[12pt]{article}
\begin{document}
%%%%%%%%%%%%%%%%%%%%%%%%%%%%%%%%%%%%%%%%%%%%%%%%%%
\def \lsim {~\mbox{${}^< \hspace*{-7pt} _\sim$}~}
\def \gsim {~\mbox{${}^> \hspace*{-7pt} _\sim$}~}
\def \leql {~\ ^< \hspace*{-7pt} _=~\ }
\def \geql {~\ ^> \hspace*{-7pt} _=~\ }
\def\preRate{\Bigl[ N \frac{\rho_\chi}{m_\chi} \Bigr]}
\def\dsdq2{\frac{d\sigma}{dq^2}}
\def\erf{\mbox{erf}}
\def\and{\mbox{~~~and~~~}}
%%%%%%%%%%%%%%%%%%%%%%%%%%%%%%%%%%%%%%%%%%%%%%%%%%%%%%%%%%%%%%%%%%%%%%%%%%%
%---- miscellaneous abrreviations
%
\newcommand{\mt}   {{\ifmmode m_{t}         \else $m_{t}$          \fi}}
\newcommand{\tb}   {{\ifmmode \tan\beta     \else $\tan\beta$      \fi}}
\newcommand{\mz}   {{\ifmmode M_{Z}         \else $M_{Z}$          \fi}}
\newcommand{\bsg}  {{\ifmmode \b\rightarrow s\gamma
                        \else $b\rightarrow s\gamma$ \fi}}
\newcommand{\Bbsg}  {{\ifmmode BR(\b\rightarrow s\gamma)
                        \else $BR(b\rightarrow s\gamma)$ \fi}}
\newcommand{\nn}{\nonumber}
\renewcommand{\floatpagefraction}{0.005}
\newcommand{\be}[1]{\begin{equation} \label{(#1)}}
\newcommand{\ee}{\end{equation}}
\newcommand{\ba}[1]{\begin{eqnarray} \label{(#1)}}
\newcommand{\ea}{\end{eqnarray}}
\newcommand{\rf}[1]{(\ref{(#1)})}
%%%%%%%%%%%%%%%%%%%%%%%%%%%%%%%%%%%%%%%%%%%%%%%%%%%%%%%%%%%%%%%%%

\begin{center}
    {\Large\bf
     Superlight neutralino as a dark matter particle candidate \\
    }

\vspace{0.3cm}

        V.A.Bednyakov$^*$\footnotemark[1],
        H.V.Klapdor-Kleingrothaus\footnotemark[2]
        and S.G.Kovalenko$^*$\footnotemark[3] \\
\smallskip
        {\it
        Max-Planck-Institut f\"{u}r Kernphysik, \\
        Postfach 103980, D-69029, Heidelberg, Germany
        }\\

\bigskip

$^*$ {\it
        Laboratory of Nuclear Problems,
	Joint Institute for Nuclear Research, \\
        Moscow region, 141980 Dubna, Russia
        }

\footnotetext[1]{E-mail: bedny@nusun.jinr.ru}
\footnotetext[2]{E-mail: klapdor@enull.mpi-hd.mpg.de}
\footnotetext[3]{E-mail: kovalen@nusun.jinr.ru}

\vspace*{0.5cm}
	{\bf Abstract}
\end{center}
\hspace*{.75in}
\vbox{
{\hsize 5in
        We address the question of how light can be the lightest
        supersymmetric particle neutralino to
        be a reliable cold dark matter (CDM) particle candidate.
        To this end we have performed a combined analysis of the parameter
	space of the Minimal Supersymmetric Standard Model (MSSM)
	taking into account cosmological and accelerator constraints
	including those from the radiative \bsg decay.
        Appropriate grand unification (GUT) scenarios were considered.

        We have found that the relaxation of gaugino mass unification
        is sufficient
	to obtain a phenomenologically and cosmologically
        viable solution of the MSSM with
        the neutralino as light as  3~GeV.

        We have found  good prospects for direct detection
        of these superlight CDM neutralinos via elastic scattering
        off various nuclei in the forthcoming experiments with
        low-threshold DM detectors.

        In a certain sense, these experiments can probe the gaugino
        mass unification giving constraints on the possible GUT
        scenarios within the MSSM.

}}

%%%%%%%%%%%%%%%%%%%%%%
\section{Introduction}
%%%%%%%%%%%%%%%%%%%%%%

	The Minimal Supersymmetric Standard Model (MSSM)
\cite{rev}
        is a leading candidate for a low energy theory
	consistent with the grand unification (GUT) idea.
        The gauge coupling constants precisely measured
        at LEP make unification in the Standard Model (SM)
        rather problematic while in the MSSM it occurs naturally
        with an excellent precision.

        The MSSM, supplied a priori with a complete set of
        the grand unification conditions, possesses a remarkable
        predictive power. 
	The complete set of the GUT conditions includes
        gauge coupling constants unification as well as unification of
        the "soft" supersymmetry (SUSY) breaking parameters at the same
        GUT scale $M_X\sim 10^{16}$ GeV. 
	Instead of this ultimate GUT scenario
        one can consider less restrictive particular GUT scenarios
        relaxing some of the GUT conditions. 
	We have no yet firm
        theoretical arguments in favor of one of these scenario.
        Analyzing the prospects for discovering SUSY in various experiments
        it is more attractive to adopt a phenomenological low-energy
        approach  and disregard certain GUT conditions.
        Following these arguments we will discuss 
        several GUT scenarios in the present paper.

        Another advantage of the MSSM is the prediction of a
        stable lightest supersymmetric particle (LSP) -- neutralino ($\chi$).
        Now the neutralino is the best known cold dark matter (CDM)
        particle candidate
\cite{roskane}.

        There is a well-known lower limit on the neutralino mass
        $m_{\chi}\geq $~18.4~GeV
\cite{LRoszk, PDB}.
	This result is strongly connected to
        the unification scenario with the universal gaugino
        masses $M_{i}(M_X) = m_{1/2}$ at the GUT scale $M_X$.
	The renormalization group evolution of $M_i$, 
	starting with  the same 
        $m_{1/2}$ value, 
	leads to tight correlation between
        the neutralino $m_{\chi_i}$ (with $i = 1-4$),
        chargino $m_{\chi_{1,2}}^\pm$  and gluino $m_{\tilde g}$ masses
        (see section 2).
        As a result, direct and indirect SUSY searches at Fermilab
        and at LEP
\cite{PDB, LEP15}
        strongly disfavor the neutralinos lighter than 18.4~GeV
        within the universal gaugino mass scenario
\cite{PDB}.
        As discussed in the following, the non-universal
        gaugino mass scenario with non-equal gaugino masses at the
        GUT scale  allows essentially lighter neutralinos.
        In a certain sense, direct searches for the
        "superlight" neutralinos in the mass range
        $m_{\chi} < $~18.4~GeV could be a test of the gaugino
        mass unification.

\smallskip

        In the present paper we will consider the discovering potential
        of DM experiments searching for the superlight DM neutralinos
        via neutralino-nucleus elastic scattering.
        Since the nuclear recoil energy in this collision is
        $E_r\sim 10^{-6} m_{\chi}$, a DM detector should
        have the low threshold $E_r^{thr}\sim$~few~KeV to be sensitive to
        the DM neutralinos as light as $m_{\chi}\sim$~few~GeV.

        There are several projects of DM experiments
        with low-threshold detectors which are able to probe
        this mass region
\cite{Bernabei}.
        Some of them are expected to run in the near future.
        These experiments will use either a new generation of cryogenic
        calorimeters
\cite{CRESST, EDELWEISS, Trofimov}
        or Germanium detectors of special configuration
\cite{Heidelberg}.

        We will argue that these experiments have good prospects
        for verification of the gaugino mass unification at the GUT scale.
        The basic reason follows from the results of our analysis of
        the MSSM parameter space within the non-universal
        gaugino mass unification scenarios.
        In these scenarios we have found superlight neutralinos 
	with masses as small as 3 GeV which produce cosmologically 
	viable relic density $0.025\leq \Omega_{\chi}\leq 1$
        and a substantial total event rate $R\sim$~1 event/kg/day
        of elastic scattering from nuclei in a DM detector.
        These values of $R$ are within the expected sensitivity
        of the above-mentioned low-threshold DM detectors.
        Therefore, the superlight DM neutralinos appearing in
        the non-universal gaugino mass GUT scenarios can be
        observed with these set-ups. Negative results of
        DM neutralino searches in the mass region
        3 GeV$\leq m_{\chi}\leq$ 18.4 GeV would discriminate
        these scenarios.

\smallskip

        The paper is organized as follows.
        In  section 2 we specify the MSSM and give formulas used
        in the subsequent sections.
        In section 3 we summarize the experimental
	and cosmological inputs for our
        analysis. Section 4 is devoted to calculation of the
        event rate of the elastic neutralino-nucleus scattering.
        In section 5 we discuss the results of our numerical analysis and
        section 6 contains the conclusion.

%%%%%%%%%%%%%%%%%%%%%%%%%%%%%%%%%%%%%%%%%%%%%%%%%%%
\section{The Minimal Supersymmetric Standard Model}
%%%%%%%%%%%%%%%%%%%%%%%%%%%%%%%%%%%%%%%%%%%%%%%%%%%

   	The  MSSM is completely specified by the standard
   	SU(3)$\times$SU(2)$\times$U(1) gauge couplings
   	as well as by the low-energy superpotential and "soft" SUSY
   	breaking terms 
\cite{rev}.
   	The most general gauge-invariant form of the R-parity
	% $(R_p = (-1)^{3B+L+2S})$
  	conserving superpotential is
\be{superpot}
{\cal W} = h_{E}L^{j}E^{c}H_{1}^{i}{\epsilon_{ij}}
    + h_{D}Q^{j}D^{c}H_{1}^{i}{\epsilon_{ij}}
    + h_{U}Q^{j}U^{c}H_{2}^{i}{\epsilon_{ij}}
    + \mu H_{1}^{i}H_{2}^{j}{\epsilon_{ij}}
\ee 
   	($\epsilon_{12}=+1$). 
	The following notations are used for
   	the quark $Q(3,2,1/6)$,  $D^{c}(\overline 3,1,1/3)$,
   	$U^{c}(\overline 3,1,-2/3)$, lepton $L(1,2,-1/2)$,
   	$E^{c}(1,1,1)$ and Higgs $H_{1}(1,2,-1/2)$, $H_{2}(1,2,1/2)$
   	chiral superfields with the SU(3)$_c\times$SU(2)$_L\times$U(1)$_Y$
   	assignment given in brackets.
   	Yukawa coupling constants $h_{E, D, U}$ are matrices in the 
	generation space, non-diagonal in the general case.
   	For simplicity we suppressed generation indices.

   	In general, the ``soft'' SUSY breaking terms are given by
\cite{soft}: 
\ba{soft} 
{\cal L}_{{SB}} &=&
- {\frac {1} {2}} {\sum_{A}} M_{A} {\bar{\lambda}_{A}}{\lambda_{A}}
      -m_{H_{1}}^{2} |H_{1}|^{2}
      -m_{H_{2}}^{2} |H_{2}|^2
            -{m^{2} _{\tilde{Q}}}  |\tilde{Q}|^2
      -{m^{2}_{\tilde{D}}} |\tilde{D^{c}}|^{2}
      - {m^{2}_{\tilde{U}}} |{\tilde{U}}^{c}|^{2} \nonumber \\
      &  -& {m^{2}_{\tilde{L}}} |\tilde{L}|^{2}
      -{m^{2}_{\tilde{E}}} |\tilde{E^{c}}|^{2}
      - (h_{E} A_{E} \tilde{L}^{j} \tilde{E}^{c} H_{1}^{i}
\epsilon_{ij}
      +h_{D} A_{D} {\tilde{Q}}^{j} {\tilde{D}}^{c} H_{1}^{i}
\epsilon_{ij} \nonumber \\
      & +& h_{U} A_{U} {\tilde{Q}}^{j} {\tilde{U}}^{c} H_{2}^{i}
\epsilon_{ij}
       +h.c)-
      (B \mu H_{1}^{i} H_{2}^{j} \epsilon_{ij}  + \mbox{h.c})
\ea%
   	As usual, $M_{3,2,1}$ are the masses of the
	SU(3)$\times$U(2)$\times$U(1) gauginos $\tilde g$, $\tilde W$, 
	$\tilde B$ and $m_i$ are the masses of scalar fields.
   	$A_L,\ A_D, \ A_U$ and $B$ are trilinear and bilinear couplings.

   	Observable quantities can be calculated in terms of the gauge and
   	the Yukawa coupling constants as well as the soft SUSY breaking
   	parameters and the Higgs mass parameter $\mu$ introduced in
Eqs.~\rf{superpot}--\rf{soft}.

   	Under the renormalization they depend on the energy scale Q
   	according to the renormalization group equations (RGE).

   	It is a common practice to implement the GUT conditions at 
	the GUT scale $M_X$.
	It allows one to reduce the  number of free parameters of the MSSM.
        A complete set of GUT conditions is:
\ba{boundary}
        m_{\tilde L}(M_X) &=& = m_{H_{1,2}}(M_X) = m_{\tilde{E}}(M_X) =
   			m_{\tilde Q}(M_X) = m_{\tilde U}(M_X)
                 		= m_{\tilde D}(M_X) = m_{0},\\
	        A_U(M_X) &=& A_D(M_X) = A_L(M_X) = A_{0}, \nn \\
%\label{(GUTtorelaxH)}        m_{H_{1,2}}(M_X) &=& m_0^{(H)}, \\
\label{(GUTtorelaxG)}        M_{i}(M_X) &=&  m_{1/2}, \\
\label{(gauge1)} \alpha_{i}(M_X) &=& \alpha_{GUT},  \ \ \
\mbox{ where  \ \ }
\alpha_1 = \frac 53 \frac{g^{\prime2}}{4\pi}, \ \
\alpha_2 = \frac{g^2}{4\pi}, \ \
\alpha_3 = \frac{g_s^2}{4\pi},
\ea
        $g'~,g$ and $g_s$ are the U(1), SU(2) and SU(3) gauge
	coupling constants.

        The above GUT conditions put very stringent constraints on the
        weak scale particle spectrum and couplings. 
	In this scenario the 
        neutralino is heavier than 18.4 GeV, as mentioned in the introduction.
        Due to specific correlations between the MSSM parameters
        at the weak scale,
        stemming from Eqs. \rf{boundary}--\rf{gauge1},  the detection rate of
        the DM neutralinos is very small, being typically 
	beyond the 
	realistic abilities of the present and future DM detectors.

        However, at present we have no strong motivation
        to impose a complete set of
        the GUT conditions in Eqs. \rf{boundary}--\rf{gauge1}
	on the MSSM.

        Therefore, in our analysis we use relaxed versions
        of the above discussed GUT
        conditions and consider two scenarios with
        the non-universal gaugino masses $M_3 = M_2\neq M_1$: \\
        {\bf (a)} with the universal
$$m_{H_{1,2}}(M_X) = m_{\tilde f}(M_X) = m_0,
        \and
$$
        {\bf (b)} non-universal
\be{scen_b}
m_{H_{1}}(M_X) = m_{H_{2}}(M_X)\neq m_0.
\ee
	GUT scale scalar masses. 
	Deviation from the GUT scale universality of
	the mass parameters in the scalar sector was previously 
	considered in Refs.
\cite{KGrLRosz}--\cite{BKK}.

        Accepting the GUT conditions, we end up with the following 
	free MSSM parameters:
        the common gauge coupling $\alpha_{GUT}$;
        the matrices of the Yukawa couplings $h_i^{ab}$,
        where $i = E, U, D$;
        the soft supersymmetry breaking parameters
        $m_0, \ m_{1/2}, \ A_0, \ B$,
        the Higgs field mixing parameter  $\mu $
        and an  additional parameter of
        the Higgs sector $m_A$ being the mass of the CP-odd neutral
        Higgs boson.
        Since the masses of the third generation are much larger
        than the masses of the first two ones,
        we consider only the Yukawa coupling of the third generation
        and drop the indices $a,b$.

        Additional constraints follow from the minimization conditions
        of the scalar Higgs potential.
        Under these conditions the bilinear coupling $B$ can be
        replaced in the given list of free parameters
        by the ratio $\tb=v_2/v_1$ of the vacuum expectation values of
        the two Higgs doublets.

        We calculate the weak-scale parameters in
Eqs.\rf{superpot}--\rf{soft}
        in terms of the above-listed free parameters on the
        basis of 2-loop RGEs following the iteration algorithm developed in
\cite{WBK}.

\smallskip

        The Higgs potential $V$ including the one-loop corrections
        $\Delta V$ can be written as:
\begin{eqnarray}
  V(H_1^0,H_2^0) &=& m^2_1|H_1^0|^2+m^2_2|H_2^0|^2-
                     m^2_3(H_1^0H_2^0+h.c.)+
\frac{g^2+g^{'2}}{8}(|H_1^0|^2-|H_2^0|^2)^2 + \Delta V, \nonumber  \\
{\rm with }\quad\Delta V&=&\frac{1}{64\pi^2}
\sum_i(-1)^{2J_i}(2J_i+1)C_im_i^4
\left[\ln\frac{m_i^2}{Q^2}-\frac{3}{2}\right],
\label{(Oneloop)}
\end{eqnarray}
   	where the sum is taken over all possible particles with the spin
	$J_i$ and with the color degrees of freedom $C_i$.
   	The mass parameters of the potential are introduced in the
	usual way as
\be{higgs_bound}
   	m_{1,2}^2 = m^2_{H_{1,2}} + \mu^2, \ \ \
        m^2_3  =  B\mu,
\ee
        They are  running parameters $m_i(Q)$  with the Q-scale dependence
        determined by the RGE.
   	The 1-loop potential
\rf{Oneloop}
	itself is Q-independent up to, irrelevant for the symmetry
	breaking, field-independent term depending on Q.
	At the minimum of this potential the neutral components of
   	the Higgs field acquire non-zero vacuum expectation values
   	$\langle H^0_{1,2}\rangle = v_{1,2}$
   	triggering the electroweak symmetry breaking with
   	$g^2(v_1^2 + v_2^2) = 2 M_W^2$.

   	The minimization conditions read
\ba{min1}
2m_1^2&=&2m_3^2\tan \beta - M_Z^2\cos 2\beta - 2\Sigma_1, \label{2m1} \\
\label{(min2)}
2m_2^2&=&2m_3^2\cot \beta + M_Z^2\cos 2\beta - 2\Sigma_2, \label{2m2}
\ea
	where
	$\Sigma_k\equiv \frac{\partial \Delta V}{\partial \psi_k}$,
	with $\psi_{1,2} = \mbox{Re} H^0_{1,2}$, \
   	are the one-loop corrections
\cite{loopewbr}:
\be{sigma1}
\Sigma_k    =  -\frac{1}{32\pi^2}
\sum_i (-1)^{2J_i}(2J_i+1)\frac{1}{\psi_k}\frac{\partial m_i^2}
   {\partial \psi_k}
   m_i^2\left(\log\frac{m_i^2}{Q^2}-1\right). 
\ee
        As a remnant of two Higgs doublets $H_{1,2}$ after the electroweak
        symmetry breaking there 
	occur five  physical Higgs particles:
        a $CP$-odd neutral Higgs boson $A$, $CP$-even neutral Higgs bosons
        $H,h$ and a pair of charged Higgses $H^{\pm}$.
        Their masses $m_A^{}$, $m_{h,H}^{}$, $m_{H^{\pm}}^{}$
        can be calculated including all 1-loop corrections
        as second derivatives of the Higgs potential in
Eq.~\rf{Oneloop}
        with respect to the corresponding fields evaluated at the minimum
\cite{erz,berz}.

   	The neutralino mass matrix  written in the basis
   	($\tilde{B}$, $\tilde{W}^{3}$, $\tilde{H}_{1}^{0}$,
	 $\tilde{H}_{2}^{0}$) has the form:
\begin{equation}
  {\cal M}_{\chi} = \left(
  \begin{array}{cccc}
    M_1 & 0   &-M_Z\cos\beta \sin_W & M_Z\sin\beta \sin_W  \\
    0   & M_2 & M_Z\cos\beta \cos_W &-M_Z\sin\beta \cos_W  \\
   -M_Z\cos\beta \sin_W & M_Z\cos\beta \cos_W & 0   & -\mu \\
    M_Z\sin\beta \sin_W &-M_Z\sin\beta \cos_W &-\mu & 0
  \end{array} \right).  
\label{(neutmix)} \\
\end{equation}
        The universal gaugino mass unification scenarios with the 
	GUT condition in 
Eq.~\rf{GUTtorelaxG}
        imply the relation
\be{gut_g}
M_1 = \frac{5}{3} \tan^2\theta_W M_2.
\ee
   	Diagonalizing the mass matrix  by virtue of the orthogonal
   	matrix ${\cal N}$ one can obtain
   	four physical neutralinos $\chi_i^{}$ with the field content
\be{admix}
\chi_i^{} = {\cal N}_{i1} \tilde{B} +  {\cal N}_{i2}  \tilde{W}^{3} +
{\cal N}_{i3} \tilde{H}_{1}^{0} + {\cal N}_{i4} \tilde{H}_{2}^{0}
\ee
   	and with masses $m_{{\chi}_i}$ being eigenvalues of
        the mass matrix 
\rf{neutmix}.
	We denote the lightest neutralino $\chi_1^{}$ by $\chi$.
	In our analysis $\chi$ is the lightest SUSY particle (LSP).

   	The chargino  mass term is
\be{charmasster}
 \left(\tilde W^-, \tilde H_1^- \right) {\cal M}_{\tilde\chi^{\pm}}
\left(  \begin{array}{c}
     \tilde W^+\\
     \tilde H_2^+ \end{array} \right)\ \  +  \ \ \mbox{h.c.}
\ee
        with the mass matrix
\be{charmix}
  {\cal M}_{\tilde\chi^{\pm}} = \left(
  \begin{array}{cc}
     M_2                  & \sqrt{2}M_W\sin\beta \\
     \sqrt{2}M_W\cos\beta & \mu
  \end{array} \right), %\label{(charmix)}
\ee
        which can be diagonalized by the transformation
\be{charginos}
	\tilde\chi^{-} = U_{i1}\tilde W^- +  U_{i2}\tilde H^-,   \ \
	\tilde\chi^{+} = V_{i1}\tilde W^+ +  V_{i2}\tilde H^+,
\ee
   	with $U^* {\cal M}_{\tilde\chi^{\pm}} V^{\dagger} =
  	\mbox{diag}(M_{\tilde\chi^{\pm}_1}, M_{\tilde\chi^{\pm}_2})$,
	where the chargino masses are
\ba{mcharginos}
 M^2_{\tilde\chi^{\pm}_{1,2}} &=& \frac{1}{2}\left[M^2_2+\mu^2+2M^2_W
 \mp \right.  \nn \\
&\mp& \left.
  \sqrt{(M^2_2-\mu^2)^2+4M^4_W\cos^22\beta +4M^2_W(M^2_2+\mu^2+2M_2\mu
  \sin 2\beta )}\right]. \nn
\ea
	It is seen from Eqs. \rf{neutmix} and \rf{charmix} 
	that in the universal gaugino mass scenarios, leading
	to relation \rf{gut_g}, the neutralino and chargino sectors
	of the MSSM are strongly correlated since they are described
	by the same set of three parameters $M_2, \mu, \tan\beta$.
	Relaxing  condition \rf{gut_g} we get one independent
	parameter in the neutralino sector, $M_1$. 
	In this case, taking place within the non-universal
	gaugino mass  GUT scenarios, the neutralino mass can be
	appreciably %essentially 
	smaller  than 18.4 GeV. 
	The latter is the lower neutralino mass bound following 
	from the experimental data
	in the case of the universal gaugino mass.

  	The mass matrices for the 3-generation sfermions
   	$\tilde t, \ \tilde b$ and $\tilde\tau$ in the 
	$\tilde f_L-\tilde f_R$ basis are:
\ba{stopmat}
{\cal M}^2_{\tilde t}&=&\left(\begin{array}{cc}
m_{\tilde Q}^2+m_t^2+\frac{1}{6}(4M_W^2-M_Z^2)\cos 2\beta &
m_t(A_t -\mu\cot \beta ) \\
m_t(A_t -\mu\cot \beta ) &
m_{\tilde U}^2+m_t^2-\frac{2}{3}(M_W^2-M_Z^2)\cos 2\beta
\end{array}  \right),
 \nonumber \\
{\cal M}^2_{\tilde b}&=&\left(\begin{array}{cc}
m_{\tilde Q}^2+m_b^2-\frac{1}{6}(2M_W^2+M_Z^2)\cos 2\beta &
m_b(A_b -\mu\tan \beta ) \\
m_b(A_b -\mu\tan \beta ) &
m_{\tilde D}^2+m_b^2+\frac{1}{3}(M_W^2-M_Z^2)\cos 2\beta
\end{array}  \right),
    \nonumber \\
{\cal M}^2_{\tilde\tau}&=&\left(\begin{array}{cc}
m_{\tilde L}^2+m_{\tau}^2-\frac{1}{2}(2M_W^2-M_Z^2)\cos 2\beta &
m_{\tau}(A_{\tau} -\mu\tan \beta ) \\
m_{\tau}(A_{\tau} -\mu\tan \beta ) &
m_{\tilde E}^2+m_{\tau}^2+(M_W^2-M_Z^2)\cos 2\beta
\end{array}  \right).               \nonumber
\ea
        For simplicity, in the sfermion mass matrices we ignored 
        non-diagonality in the generation space, which is important
        only for the \bsg-decay.

%%%%%%%%%%%%%%%%%%%%%%%%%%%%%%%%%%%%%%%%%%%%%%%
\section{The Constrained MSSM parameter space}
%%%%%%%%%%%%%%%%%%%%%%%%%%%%%%%%%%%%%%%%%%%%%%%

        In this section we summarize the theoretical and
	experimental constraints used in our analysis.

        The solution of the gauge coupling constants unification (see
Eq.~\rf{gauge1}) allows us to define the unification scale $M_X$. 
	Our numerical procedure is based on the 2-loop RGEs.
        We use the world averaged values of the gauge
        coup\-lings  at the $Z^0$ energy obtained from a fit to the 
	LEP data
\cite{LEP},
        $M_W$ \cite{PDB} and \mt\cite{CDF,D0}:
\ba{world}
  \alpha^{-1}(M_Z)             & = & 128.0\pm0.1,\\
  \sin^2\theta_{\overline{MS}} & = & 0.2319\pm0.0004,\\
  \alpha_3                     & = & 0.125\pm0.005.
\ea
        The value of the fine structure constant $\alpha^{-1}(M_Z)$
	was updated from
\cite{dfs} by using new data on the hadronic vacuum polarization
\cite{EJ95}.
        The standard relations are implied:
\be{standard}
\alpha_1 = \frac{5}{3}\frac{\alpha}{\cos^2\theta_W}, \ \ \
       \alpha_2 = \frac{\alpha}{\sin^2\theta_W}.
\ee
	SUSY particles have not been found so far and from the searches
	at LEP one knows that the lower limit on the charged sleptons
	is half the $Z^0$ ~mass (45 GeV)
~\cite{PDB} and the Higgs mass has
	to be above 60 GeV
~\cite{higgslim,sopczak},
	while the sneutrinos have to be above 41 GeV
~\cite{PDB}.
        For the charginos the preliminary lower limit of 65 GeV was obtained
        from the LEP 140~GeV  run
~\cite{LEP15}.
        The above mass limits are incorporated in our analysis.

        Radiative corrections trigger spontaneous symmetry breaking in the
        electroweak sector.
        In this case the Higgs potential has its minimum for the non-zero
        vacuum expectation values of the Higgs fields $H^0_{1,2}$.
        Solving $\mz$ from
Eqs.~\rf{min1}--\rf{min2} yields:
\be{defmz}
\frac{\mz^2}{2}=\frac{m_1^2+\Sigma _1-(m_2^2+\Sigma _2) \tan^2\beta}
{\tan^2\beta-1},
\ee
        where $\Sigma_1$ and $\Sigma_2$ are defined in
Eq.~\rf{sigma1}. 
        This is an important constraint which relates the true vacuum
        to the physical Z-boson mass $M_Z = 91.187\pm 0.007$~GeV.

        Another stringent constraint is imposed by the branching ratio
        \Bbsg, measured by the CLEO collaboration
\cite{cleo94}
        to be $BR(b\to s \gamma)= (2.32\pm0.67)\times10^{-4}$.

        In the MSSM this flavor changing neutral current (FCNC)
        process receives contributions from $H^\pm-t$,
        $\tilde{\chi}^\pm - \tilde{t}$ and $\tilde{g}-\tilde{q}$ loops
        in addition to the SM\ \ $W-t$ loop.
        The ${\chi} -\tilde{t}$ loops, which are expected to be much
        smaller, have been neglected
\cite{BerBorMasRi,borz}.
        The $\tilde{g}-\tilde{q}$ loops are proportional to \tb.
        It was found
\cite{WBK}
        that this contribution should  be small,
        even in the case of large \tb and therefore it can be neglected.
        The chargino contribution, which becomes large for large \tb and
        small chargino masses, depends sensitively on the splitting
        of the two stop mass eigenstates $\tilde t_{1,2}$.

	Within the MSSM the following ratio has been calculated
\cite{BerBorMasRi}:
$$ 
	\frac{BR(b\to s\gamma)}{BR(b\to c e
	\bar{\nu})}=\frac{|V_{ts}^*V_{tb}|^2}{|V_{cb}|^2}
	K_{NLO}^{QCD} \frac{6\alpha}{\pi}
	\frac{\left[\eta^{16/23}A_\gamma+\frac{8}{3}(\eta^{14/23}
	-\eta^{16/23})A_g+C\right]^2}{I(x_{cb})
	[1-(2/3\pi)\alpha_s(m_b)f(x_{cb})]}, 
$$ 	where
$$ 	C \approx 0.175, \ \ I(x_{cb}) = 0.4847, \ \
   	\eta=\alpha_s(M_W)/\alpha_s(m_b), \ \ f(x_{cb})=2.41.
$$
        Here $f(x_{cb})$ represents corrections from leading order QCD
        to the known semileptonic $b\to c e \bar{\nu}$ decay rate;
        $I(x_{cb})$  is a phase space factor;
        $x_{cb} = m_c/m_b=0.316$; 
	$K_{NLO}^{QCD}$ describes the next-to-leading-order QCD corrections
\cite{alibsg}.
        $A_{\gamma,g}$ are the coefficients of the effective operators
        for $bs$-$\gamma$ and  $bs$-$g$ interactions respectively;
        $C$ describes mixing with the four-quark operators.
	The ratio of CKM matrix elements
	$\frac{|V_{ts}^*V_{tb}|^2}{|V_{cb}|^2}=0.95$ was taken from
\cite{burasb}.

\bigskip

   	Assuming that the neutralinos  form a dominant part of
   	the DM in the universe one obtains a cosmological constraint
   	on the neutralino relic density.

   	The present lifetime of the universe is at least $10^{10}$ years,
   	which implies an upper limit on the expansion rate and
   	correspondingly on the total relic abundance.
   	Assuming $h_0>0.4$ one finds that  the contribution of
   	each relic particle species $\chi$  has to obey
\cite{kolb}:
\be{Age_Uni}
 	\Omega_\chi h^2_0<1,
\ee
   	where the relic density parameter  $\Omega_\chi = \rho_\chi/\rho_c$
   	is the ratio of the relic neutralino mass  density $\rho_\chi$
   	to  the critical one
   	$\rho_c = 1.88\cdot 10^{-29}$h$^2_0$g$\cdot$cm$^{-3}$.

   	We calculate $\Omega_{\chi} h^2_0$  following the standard
   	procedure on the basis of the approximate formula
\ba{omega}
\Omega_{\chi} h^2_0 &=& 2.13\times 10^{-11}
	\left(\frac{T_{\chi}}{T_{\gamma}}\right)^3
	\left(\frac{T_{\gamma}}{2.7K^o}\right)^3
	\times N_F^{1/2}
	\left(\frac{{\mbox{GeV}}^{-2}}{a x_F + b x_F^2/2}\right).
\ea
   	Here $T_{\gamma}$ is the present day photon temperature,
   	$T_{\chi}/T_{\gamma}$ is the reheating factor,
   	$x_F = T_F/m_{\chi} \approx 1/20$, $T_F$ is
   	the neutralino freeze-out temperature, and $N_F$ is the total
   	number of degrees of freedom at $T_F$.
	The coefficients $a, b$ are determined from the
	non-relativistic expansion
\be{expan}
<\sigma_{annyh} v> \approx a + b x
\ee
   	of the thermally averaged cross section of neutralino annihilation.
        We adopt an approximate treatment ignoring
        complications, which appear when the expansion
\rf{expan} fails
\cite{omega}.
        In our analysis all possible channels of the $\chi-\chi$ annihilation
        are taken into account.
	The complete list of the formulas,
	which we used, for the coefficients  $a, b$ and numerical values
   	for the other parameters in Eqs.
\rf{omega} and \rf{expan}
   	can be found in
\cite{drno}.

   	Since the neutralinos are mixtures of gauginos and
        higgsinos, the annihilation can occur via
        s-channel exchange of the $Z^0$, Higgs bosons and
   	t-channel exchange of a scalar particle, like a selectron
\cite{relic}.
        Therefore, the cosmological constraint in Eq. \rf{Age_Uni}
        substantially %essentially 
	restricts the MSSM parameter space, as discussed
        by many groups
\cite{roskane,drno,rosdm,relictst}.

        In the analysis we ignore possible rescaling of the local
        neutralino density $\rho$ which may occur in the region of the
        MSSM parameter space where $\Omega_\chi h^2< 0.025$
\cite{Gelm,Bot}.
        At lower relic densities DM neutralinos cannot saturate even
        galactic halos in the universe and the  presence of additional
        DM components should be taken into account.
        One may assume that it can be done by virtue of the above-mentioned
        rescaling ansatz.
        Let us note that the halo density is a very uncertain quantity.
        Its actual value can be one order of magnitude smaller (or larger)
        than the quoted value 0.025
\cite{GonBer}.
        The SUSY solution  of the DM problem with
	such low neutralino density becomes questionable.
        Therefore, we simply skip the corresponding domains
        of the MSSM parameter space as cosmologically
        uninteresting.

        Thus, we assume neutralinos to be a dominant component of
        the DM halo of our galaxy with a density
        $\rho_{\chi}$ = 0.3 GeV$\cdot$cm$^{-3}$ in the solar vicinity.

%%%%%%%%%%%%%%%%%%%%%%%%%%%%%%%%%%%%%%%%%%%%%%%%%%
\section{Neutralino-Nucleus Elastic Scattering}
%%%%%%%%%%%%%%%%%%%%%%%%%%%%%%%%%%%%%%%%%%%%%%%%%%

        A dark matter event is elastic scattering of a DM neutralino from
   	a target nucleus producing a nuclear recoil which can be
	detected by a detector.
        The corresponding event rate depends on the distribution of
        the DM neutralinos in the solar vicinity and
        the cross section $\sigma_{el}(\chi A)$ of  neutralino-nucleus
        elastic scattering.
        In order to calculate $\sigma_{el}(\chi A)$  one should specify
        neutralino-quark interactions.
        The relevant low-energy effective Lagrangian can be written
        in a general form as
\be{Lagr}
  L_{eff} = \sum_{q}^{}\left( {\cal A}_{q}\cdot
      \bar\chi\gamma_\mu\gamma_5\chi\cdot
                \bar q\gamma^\mu\gamma_5 q +
    \frac{m_q}{M_{W}} \cdot{\cal C}_{q}\cdot\bar\chi\chi\cdot\bar q q\right)
      \ +\ O\left(\frac{1}{m_{\tilde q}^4}\right),
\ee
        where terms with vector and pseudoscalar quark currents are
        omitted being negligible in the case of non-relativistic
        DM neutralinos with typical velocities $v_\chi\approx 10^{-3} c$.

        In the Lagrangian
\rf{Lagr} we also neglect terms which appear in supersymmetric models
	at the order of $1/m_{\tilde q}^4$ and higher,
	where $m_{\tilde q}$ is the mass of the scalar superpartner
	$\tilde q$ of the quark $q$.
        These terms, as pointed out in Ref.
\cite{drno},
        are potentially
        important in the spin-independent neutralino-nucleon scattering,
  	especially in the domains of the MSSM  parameter space where
  	$m_{\tilde q}$ is close to the neutralino mass $m_{\chi}$.
  	Below we adopt the approximate treatment of these terms
        proposed in
\cite{drno}, which allows "effectively"
        absorbing them into the coefficients ${\cal C}_q$ in a wide region
  	of the SUSY model parameter space.
        Our formulas for the coefficients ${\cal A}_q$ and ${\cal C}_q$
        of the effective Lagrangian take into account squark mixing
   	$\tilde{q}_L-\tilde{q}_R$ and the contribution of
   	both CP-even Higgs bosons $h, H$:
\ba{Aq1}
 {\cal A}_{q} =
	&-&\frac{g^{2}}{4M_{W}^{2}}
	   \Bigl[\frac{{\cal N}_{14}^2-{\cal N}_{13}^2}{2}T_3 \nn \\
        &-& \frac{M_{W}^2}{m^{2}_{\tilde{q}1} - (m_\chi + m_q)^2}
	   (\cos^{2}\theta_{q}\ \phi_{qL}^2
	   + \sin^{2}\theta_{q}\ \phi_{qR}^2) \nn \\
        &-& \frac{M_{W}^2}{m^{2}_{\tilde{q}2} - (m_\chi + m_q)^2}
	     (\sin^{2}\theta_{q}\ \phi_{qL}^2
	     + \cos^{2}\theta_{q}\ \phi_{qR}^2) \nn \\
        &-& \frac{m_{q}^{2}}{4}P_{q}^{2}\left(\frac{1}{m^{2}_{\tilde{q}1}
		- (m_\chi + m_q)^2}
             + \frac{1}{m^{2}_{\tilde{q}2}
                - (m_\chi + m_q)^2}\right) \nn \\
        &-& \frac{m_{q}}{2}\  M_{W}\  P_{q}\  \sin2\theta_{q}\
            T_3 ({\cal N}_{12} - \tan\theta_W {\cal N}_{11}) \nn \\
	&\times&\left( \frac{1}{m^{2}_{\tilde{q}1}- (m_\chi + m_q)^2}
    - \frac{1}{m^{2}_{\tilde{q}2} - (m_\chi + m_q)^2}\right)\Bigr],
% \\[0.5cm]
\ea
%%%%%%%%
\ba{Cq1}
 {\cal C}_{q} =
	&-&  \frac{g^2}{4} \Bigl[\frac{F_h}{m^2_{h}} h_q +
			\frac{F_H}{m^2_{H}} H_q \nn \\
	&+& P_q \left(\frac{\cos^{2}\theta_{q}\ \phi_{qL} -
     \sin^{2}\theta_{q}\ \phi_{qR}}{m^{2}_{\tilde{q}1} - (m_\chi + m_q)^2}
         -\frac{\cos^{2}\theta_{q}\ \phi_{qR} -
     \sin^{2}\theta_{q}\ \phi_{qL}}{m^{2}_{\tilde{q}2} - (m_\chi +
	m_q)^2}\right) \nn \\
	&+& \sin2\theta_{q}(\frac{m_q}{4 M_W} P_{q}^{2} -
                           \frac{M_W}{m_q} \phi_{qL}\ \phi_{qR}) \nn \\
	&\times&\left(\frac{1}{m^{2}_{\tilde{q}1} - (m_\chi + m_q)^2} -
\frac{1}{m^{2}_{\tilde{q}2} - (m_\chi + m_q)^2}\right)\Bigr].
\ea
%%%%%%%
     Here
\ba{Fq1}
        F_{h} &=& ({\cal N}_{12} - {\cal N}_{11}\tan\theta_W)
        ({\cal N}_{14}\cos\alpha_H + {\cal N}_{13}\sin\alpha_H), \nn \\
        F_{H} &=& ({\cal N}_{12} - {\cal N}_{11}\tan\theta_W)
        ({\cal N}_{14}\sin\alpha_H - {\cal N}_{13}\cos\alpha_H),\nn \\
h_q &=&\bigl(\frac{1}{2}+T_3\bigr)\frac{\cos\alpha_H}{\sin\beta}
          - \bigl(\frac{1}{2}-T_3\bigr)\frac{\sin\alpha_H}{\cos\beta},\nn \\
H_q &=& \bigl(\frac{1}{2}+T_3\bigr)\frac{\sin\alpha_H}{\sin\beta}
      + \bigl(\frac{1}{2}-T_3\bigr)\frac{\cos\alpha_H}{\cos\beta}, \nn \\
\phi_{qL} &=& {\cal N}_{12} T_3 + {\cal N}_{11}(Q -T_3)\tan\theta_{W},\nn \\
\phi_{qR} &=& \tan\theta_{W}\  Q\  {\cal N}_{11}, \nn \\
P_{q} &=&  \bigl(\frac{1}{2}+T_3\bigr) \frac{{\cal N}_{14}}{\sin\beta}
          + \bigl(\frac{1}{2}-T_3\bigr) \frac{{\cal N}_{13}}{\cos\beta}. \nn
\ea
        The above formulas coincide with the relevant formulas
        in  \cite{drno} neglecting the terms $\sim 1/m_{\tilde q}^4$
   	and higher.
        These terms are taken into account ``effectively'' by
   	introducing an ``effective''  stop quark $\tilde t$
        propagator.

        A general representation of the differential cross section
        of neutralino-nucleus scattering can be given in terms of
        three spin-dependent ${\cal  F}_{ij}(q^2)$ and
        one spin-independent ${\cal F}_{S}(q^2)$ form factors
\cite{EV}
\be{cs}
\dsdq2(v,q^2)=\frac{8 G_F}{v^2} \left(
   a_0^2\cdot {\cal F}_{00}^2(q^2) +
   a_0 a_1 \cdot {\cal F}_{10}^2(q^2) +
   a_1^2\cdot {\cal F}_{11}^2(q^2)
   + c_0^2\cdot A^2\ {\cal F}_{S}^2(q^2)
   \right),
\ee
        where  $v$ is a projectile neutralino velocity and $q$ 
	is the momentum transferred to the nucleus.
   	The last term corresponding to the spin-independent scalar 
	interaction gains coherent enhancement $A^2$  
	($A$ is the atomic weight 
	of a nucleus in the reaction).
   	The coefficients $a_{0,1}, c_0$ do not depend on the nuclear structure
   	and relate to the parameters ${\cal A}_q, 
	{\cal C}_q$ of the effective Lagrangian 
\rf{Lagr} as well as to the parameters $\Delta q, f_s, \hat{f}$
   	characterizing the nucleon structure. One has the relationships
\ba{rel1}
a_0 &=& ( {\cal A}_u + {\cal A}_d)
        ( \Delta u + \Delta d ) + 2 \Delta s {\cal A}_s, \nn \\
a_1 &=& ({\cal A}_u - {\cal A}_d)
        (\Delta u - \Delta d ),  \nn \\
c_0  &=& \hat f \frac{m_u {\cal C}_{u}
        + m_d {\cal C}_{d}}{m_u + m_d}
+ f_{s} {\cal C}_{s}  + \frac{2}{27}(1- f_{s} - \hat f)({\cal C}_{c}
        + {\cal C}_{b} + {\cal C}_{t}).
\ea
   Here  $\Delta q^{p(n)}$ are
        the fractions of the proton (neutron) spin carried by the quark $q$.
        The standard definition is
\be{Spin}
   <p(n)|\bar q\gamma^\mu\gamma_5 q|p(n)> = 2 S_{p(n)}^{\mu} \Delta q^{p(n)},
\ee
       	where $S_{p(n)}^{\mu}=(0,\vec{S}_{p(n)})$ is the 4-spin
	of the nucleon.
        The parameters $\Delta q^{p(n)}$ can be extracted from the data on
	polarized nucleon structure functions
\cite{EMC,SMC2} and the hyperon semileptonic decay data
\cite{Dqextract}.

	We use  $\Delta q$ values extracted both from the EMC
\cite{EMC} and SMC
\cite{SMC2} data.
        The other nuclear structure parameters $f_s$ and $\hat f$
        in formula  \rf{rel1} are defined as follows:
\ba{Scal}
   <p(n)|(m_{u} + m_{d})(\bar{u}u + \bar{d}d)|p(n)> &=& 2\hat f M_{p(n)}
           \bar \Psi \Psi, \\
\nn
   <p(n)|m_{s}\bar{s}s|p(n)> &=& f_{s} M_{p(n)}\bar \Psi \Psi.
\ea
        The values extracted from the data under certain theoretical
        assumptions are
\cite{ChengGasser}:
        \ba{f}
   \hat{f} = 0.05\ \ \ \ \ \ \mbox{and}\ \ \ \ \ \ f_{s} = 0.14.
        \ea
   	The strange quark contribution $f_{s}$ is known to be uncertain
        to about a factor of \  2.
	Therefore we take its value 
        within the interval $ 0.07 < f_{s} < 0.3$
\cite{ChengGasser,Hatsuda}.

        The nuclear structure comes into play via the form factors
        ${\cal F}_{ij}(q^2), {\cal F}_{S}(q^2)$ in Eq. \rf{cs}.
        The spin-independent
        form factor ${\cal F}_{S}(q^2)$ can be represented as the normalized
        Fourier transform of a spherical nuclear ground state density
   	distribution $\rho({\bf r})$.
   	We use the standard Woods-Saxon inspired distribution
\cite{Engel}.
   	It leads to the form factor
\be{fourier}
{\cal F}_{S}(q^2) = \int d^3{\bf r} \rho({\bf r}) e^{i {\bf r q}}
= 3\frac{j_1(q R_0)}{q R_0} e^{-\frac{1}{2} (qs)^2},
\ee
   	where $R_0 = (R^2 - 5 s^2)^{1/2}$ and $s \approx 1$ fm are the
   	radius and the thickness of a spherical nuclear surface,
   	$j_1$ is the spherical Bessel function of index 1.

   	Spin-dependent form factors ${\cal F}_{ij}(q^2)$ are much more
   	nuclear-model-dependent quantities.
        In the last few years noticeable progress
        in detailed nuclear-model calculations of these form factors
        has been achieved.
   	For many nuclei of interest in the DM search  they have been 
	calculated within various models of nuclear structure
\cite{Ressell},
\cite{Nikolaev}.
        Unfortunately, these calculations do not cover all isotopes
        which we are going to consider in the present paper.
        Therefore, we use a  simple parameterization of the
        $q^2$ dependence of
   	${\cal F}_{ij}(q^2)$ in the form of a Gaussian with the
        {\em rms} spin radius
   	of the nucleus calculated in the harmonic well potential
\cite{EF}.
   	For our purposes this semi-empirical scheme is sufficient.

        An experimentally observable quantity is the differential event rate
        per unit mass of the target material
\be{drate1}
\frac{dR}{dE_r} = \preRate
    \int^{v_{max}}_{v_{min}} dv f(v) v \dsdq2 (v, E_r),
\ee
        where $E_r$ is the nuclear recoil energy.
        The function $f(v)$ is the velocity distribution of neutralinos
        in the earth's frame. In the galactic frame it is usually assumed
        to have an approximate Maxwellian form.
	$v_{max} = v_{esc} \approx$ 600 km/s and
	$\rho_{\chi}$ = 0.3 GeV$\cdot$cm$^{-3}$  are the escape velocity
	and the mass density of the relic neutralinos in
   	the solar vicinity;
$v_{min} = \left(M_A E_r/2 M_{red}^2\right)^{1/2}$ with
$M_A$ and $M_{red}$ being the mass of nucleus $A$ and the reduced
   	mass of the neutralino-nucleus system, respectively.
	Note that $ q^2 = 2 M_A E_r$.

   	The differential event rate is the most appropriate
	quantity for comparing with
        the observed recoil spectrum. It allows one to take properly
        into account the spectral characteristics of a specific DM detector
        and to discriminate a background.
        However, discussing general problems and prospects of DM detection
        it is enough to consider the total  event rate $R$\
        integrated over the whole kinematical domain of recoil energy.
        Notice that this quantity is less sensitive
        to details of the nuclear structure than
        the differential event rate in Eq.
\rf{drate1}.
   	The $q^2$ shape of the form factors ${\cal F}_{ij}(q^2),
   	{\cal F}_{S}(q^2)$ in Eq.~\rf{cs} may 
	substantially %essentially 
	change from one nuclear model to another.
   	Integration over $q^2$, as in the case of
   	the total event rate $R$, reduces this model dependence.

        The present paper aims at the general
        investigation of detectability of the cosmologically
        viable "superlight" DM neutralinos independently of
        a specific DM detector.
        Therefore, we may use the total event rate $R$ as a characteristic
        of the DM signal.

%%%%%%%%%%%%%%%%%%%%%%%%%%%%%%%%%%%%%%%%%%%
\section{Numerical Analysis and Discussion}
%%%%%%%%%%%%%%%%%%%%%%%%%%%%%%%%%%%%%%%%%%%

	In our numerical analysis we randomly scanned the MSSM
	parameter space including all experimental and cosmological
        constraints discussed in  section 3.
        Two GUT scenarios with the non-universal gaugino mass
        $M_3 = M_2\neq M_1$ (see 
Eqs. \rf{boundary}--\rf{gauge1}
        and the comments below them) have been considered:
        {\bf (a)} with the universal
        $m_{H_{1,2}}(M_X) = m_{\tilde f}(M_X) = m_0$ and
        {\bf (b)} non-universal
        $m_{H_{1}}(M_X) = m_{H_{2}}(M_X)\neq m_0$
        GUT scale scalar masses.

        The free parameters of both scenarios and intervals of their
        variation in our analysis are given in table I.

%%%%%%%%%%%%%%%%%%%%%%%%%%%%%%%%%%%%%%%%%%%%%%%%%%%%%%%%%%%%%%
\begin{table}[t]
\vspace*{-0.5cm}
\begin{center}
\begin{tabular}{|c|c|cc|} \hline
scenario&{\small  Parameter} & \multicolumn{2}{c|}{\small Lower and Upper Bounds } \\ \hline
a,b &   $\tan\!\beta$  &     1        &    50      \\ %\hline
a,b &   $ m_0$         &     0~TeV    &    1~TeV   \\ %\hline
a,b &   $ A_0$         &    -1~TeV    &    1~TeV      \\ %\hline
a,b &   $ \mu$         &    -1~TeV    &    1~TeV   \\ %\hline
a,b &   $ M_2$         &     0~TeV    &    1~TeV   \\ %\hline
a,b &   $ M_1$         &   -1~TeV    &    1~TeV   \\ %\hline
b   &   $ m_A$         &   50~GeV    &    1~TeV   \\ \hline
\end{tabular}
\end{center}
\hspace*{0.8in}
\vbox{
{\hsize 5in
        TABLE I. 
	The MSSM parameters within the GUT scenarios (a), (b)
        and intervals of their variations in the present numerical analysis.

}}

\end{table}
%%%%%%%%%%%%%%%%%%%%%%%%%%%%%%%%%%%%%%%%%%%%%%%%%%%%%%%%%%%%%%%%%%%%
% 1 %%%%%%%%%%%%%%%%%%%%%%%%%%%%%%%%%%%%%%%%%%
\begin{figure}[ht]
\begin{picture}(5,4.5)
\put(0,0.0){\includegraphics{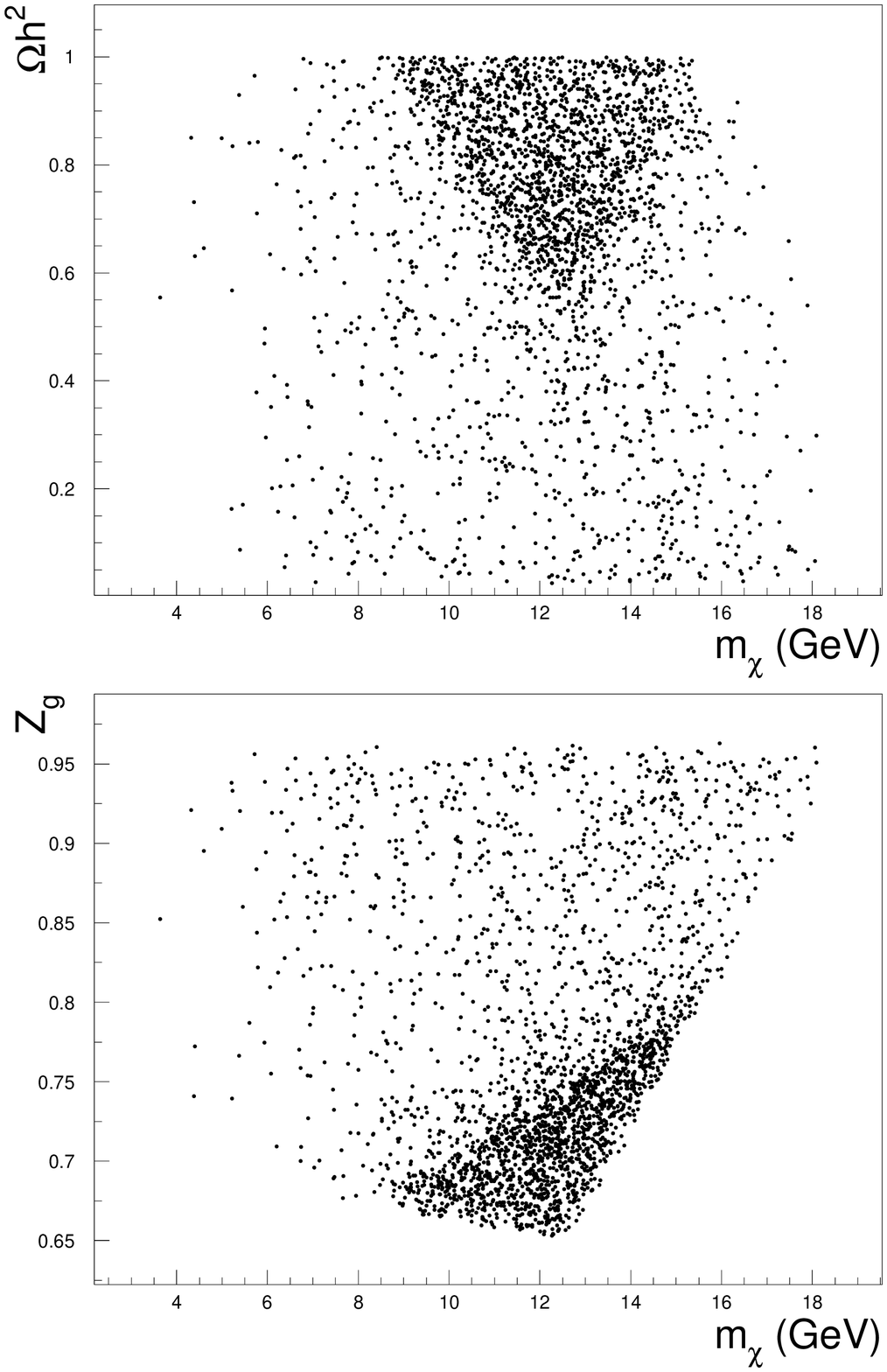}}
\end{picture}
       
	FIG. 1.
	The neutralino relic density $\Omega h^2$ and the gaugino 
	fraction $Z_g={\cal N}_{11}^2 + {\cal N}_{12}^2$
	versus the neutralino mass $m_{\chi}$ in GUT scenario (a).
\end{figure}
%%%%%%%%%%%%%%%%%%%%%%%%%%%%%%%%%%%%%%%%%%%%%%%%%%%%%%%%%%%
% 2 %%%%%%%%%%%%%%%%%%%%%%%%%%%%%%%%%%%%%%%%%%%%
\begin{figure}[t]
\begin{picture}(5,4)
\put(0,0.0){\includegraphics{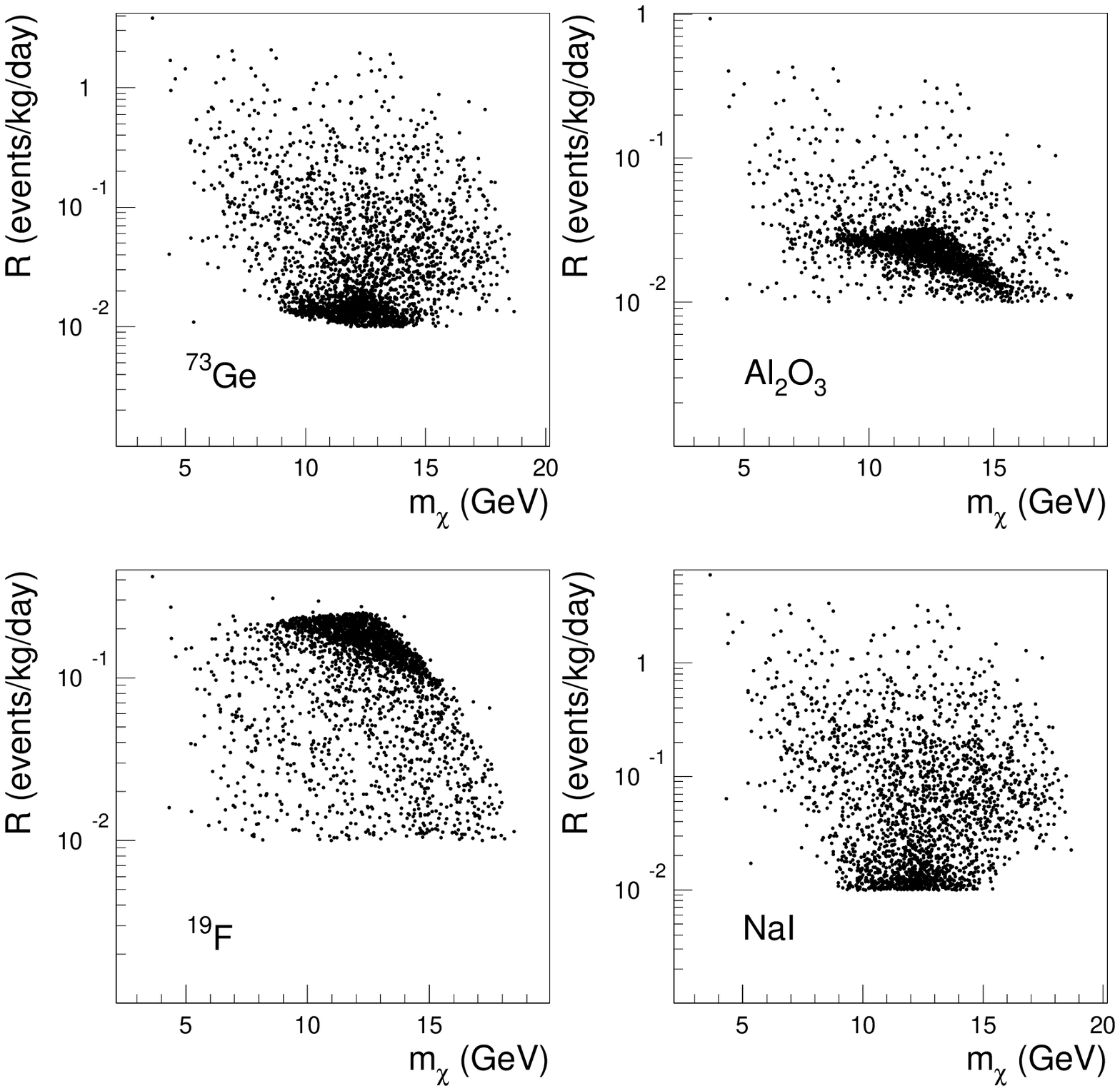}}
\end{picture}

        FIG. 2.
        The expected DM neutralino event rate $R$\
        versus the neutralino mass $m_\chi$
        for various isotopes of experimental interest.
        GUT scenario (a).
        All points with $R\leq$ 0.01 events/kg/day
        are cut off.

\end{figure}
%%%%%%%%%%%%%%%%%%%%%%%%%%%%%%%%%%%%%%%%%%%%%%%%

        After the scan procedure had been finished we found the 
	superlight neutralinos in the mass range
        3 GeV $\leq m_{\chi}\leq$ 18.4 GeV in both GUT scenarios (a)
	and (b). 
	The criterion we used to stop the running numerical 
	code was  stabilization of the $m_{\chi}$ lower bound.

        Recall that 18.4 GeV is the widely cited lower neutralino mass bound
        valid within the complete GUT scenario with
        relations  
\rf{boundary}--\rf{gauge1} at the GUT scale.
        Scenario (a) implies a minimal relaxation of the complete
        GUT scenario necessary to obtain the superlight neutralinos.
        Scenario (b) does not change the lower $m_{\chi}$ bound 3 GeV, but
        allows a much larger DM event rate $R$ than scenario (a).
        We have calculated the total event rate for
	germanium  ($^{73}$Ge),
        sapphire  (Al$_2$O$_3$),
	fluorine ($^{19}$F) and
        sodium iodide (NaI).

\smallskip

        The main results of our analysis are presented in
        Figs.~1--5 in the form of scatter plots.
        In Figs.~2--4 we cut off all points with $R\leq$ 0.01 events/kg/day
        since they lie beyond the sensitivity of the present and
        the near-future DM detectors.

        Figure 1 shows the relic density, $\Omega_\chi h^2$,
        produced by the superlight neutralinos and their gaugino fraction,
        $Z_g={\cal N}_{11}^2 + {\cal N}_{12}^2$ (see 
Eq. \rf{admix}).
        It is seen that the majority of the points are concentrated
        in the region of a large relic density, especially within the
        interval 8 GeV $\leq m_{\chi}\leq$ 16 GeV. 
	However, the neutralinos with a marginal mass of 3 GeV can 
	also produce  cosmologically interesting values of  
	$\Omega_\chi h^2$.
        Therefore we conclude that the superlight neutralinos can comprise
        a dominant part of the CDM. 
	The field composition of these neutralinos characterized by 
	the gaugino fraction $Z_g$ shows specific tendencies. 
	A large domain of the MSSM parameter space 
	(the volume is proportional to the number of points in the scatter
        plots) corresponds to the mixed $Z_g\sim 0.7$ gaugino-higgsino states
        around the mass value 12 GeV. 
	For larger $m_{\chi}$ neutralinos lose %are loosing 
	their higgsino component and become mostly gauginos.

% 3 %%%%%%%%%%%%%%%%%%%%%%%%%%%%%%%%%%%%%%%%%%%%
\begin{figure}[t]
\begin{picture}(5,4)
\put(0,0.0){\includegraphics{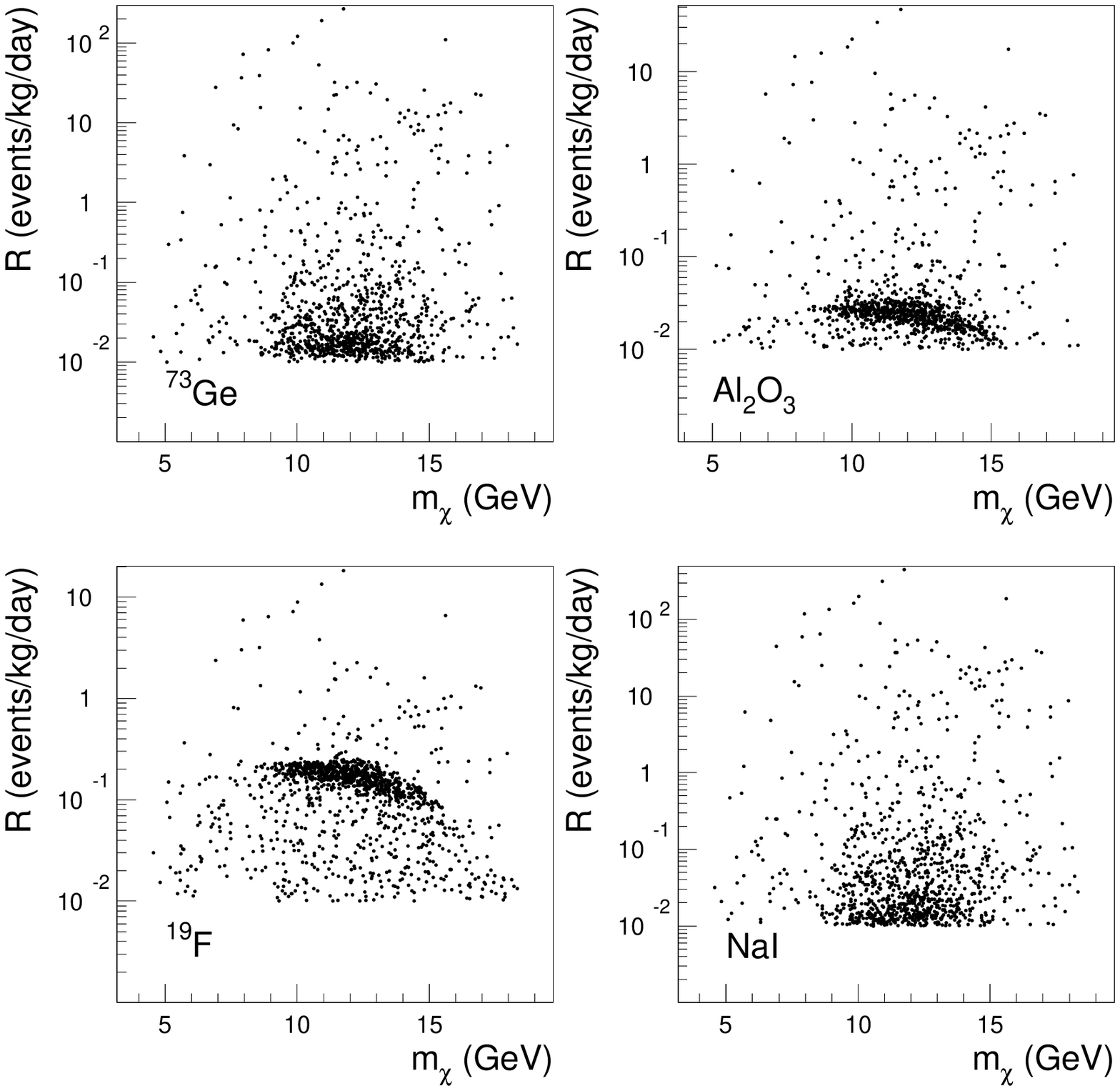}}
\end{picture}

        FIG. 3.
        The same as in Fig. 2 but for  large 
	event rate GUT scenario (b).

\end{figure}
%%%%%%%%%%%%%%%%%%%%%%%%%%%%%%%%%%%%%%%%%%%%%%%%%%%%%%%%%%%

        In Figures~2--3 we present the calculated event rate, $R$, for
        $^{73}$Ge, Al$_2$O$_3$, $^{19}$F and NaI versus the neutralino
        mass $m_\chi$ in  two GUT scenarios (a) and (b).
        These figures demonstrate that in both scenarios there are
        a lot of points within the reach of the near-future
        low-threshold DM detectors mentioned in the introduction.
        Their sensitivities are expected to be
        at the level of $R\geq 0.1$ events/kg/day.
        In the scenario (b) points extend up to
        $R\sim 300\div500$ events/kg/day for Ge, NaI and
        $R\sim 20\div50$ events/kg/day for $^{19}$F, Al$_2$O$_3$.
        In  scenario (a) these values are reduced
        by approximately a factor of 10.

        In unification scenario (b) with non-universal
        Higgs mass parameters (see 
Eq. \rf{scen_b})
        the Higgs and sfermion masses are not strongly correlated.
        As discussed in
\cite{BKK}, this relaxation of the complete unification
        in the scalar sector makes it possible 
	to avoid one of the most stringent
        theoretical limitations on the allowed values of the DM neutralino
        event rate.
        In this case the masses of the CP-even and charged Higgses
        $m_{H,h}$, $m_{H^\pm}$ are calculated in terms of the
        CP-odd Higgs mass, $m_A$, as an extra free parameter.

% 4 %%%%%%%%%%%%%%%%%%%%%%%%%%%%%%%%%%%%%%%%%%
\begin{figure}[t]
\begin{picture}(5,4)
\put(0,0.0){\includegraphics{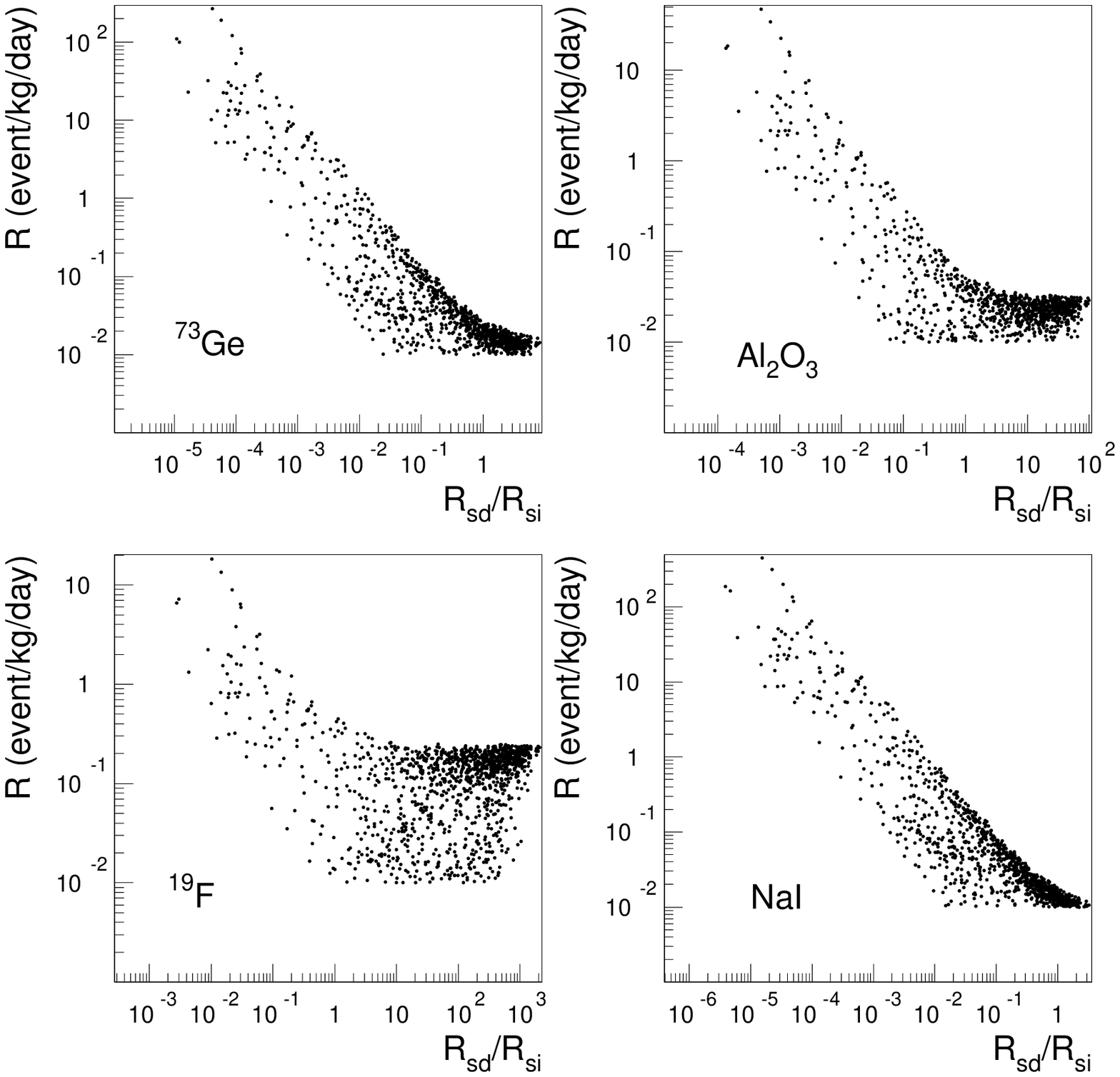}}
\end{picture}

        FIG. 4.
        Spin sensitivities of various isotopes of experimental interest.
        Given is the total event rate $R$ in GUT scenario (b)
        versus the ratio $R_{sd}/R_{si}$.  $R_{sd}$ and $R_{si}$ are
        the spin-dependent and spin-independent components of $R$
        ($R=R_{sd}+R_{si}$).
        All points with $R\leq$ 0.01 events/kg/day are cut off.

\end{figure} %  voffset=100 -> Up
%%%%%%%%%%%%%%%%%%%%%%%%%%%%%%%%%%%%%%%%%%%%%%%%%%%%%%%%%%%
% 5 %%%%%%%%%%%%%%%%%%%%%%%%%%%%%%%%%%%%%%%%%%
\begin{figure}[t]
\begin{picture}(5,4)
\put(0,0.0){\includegraphics{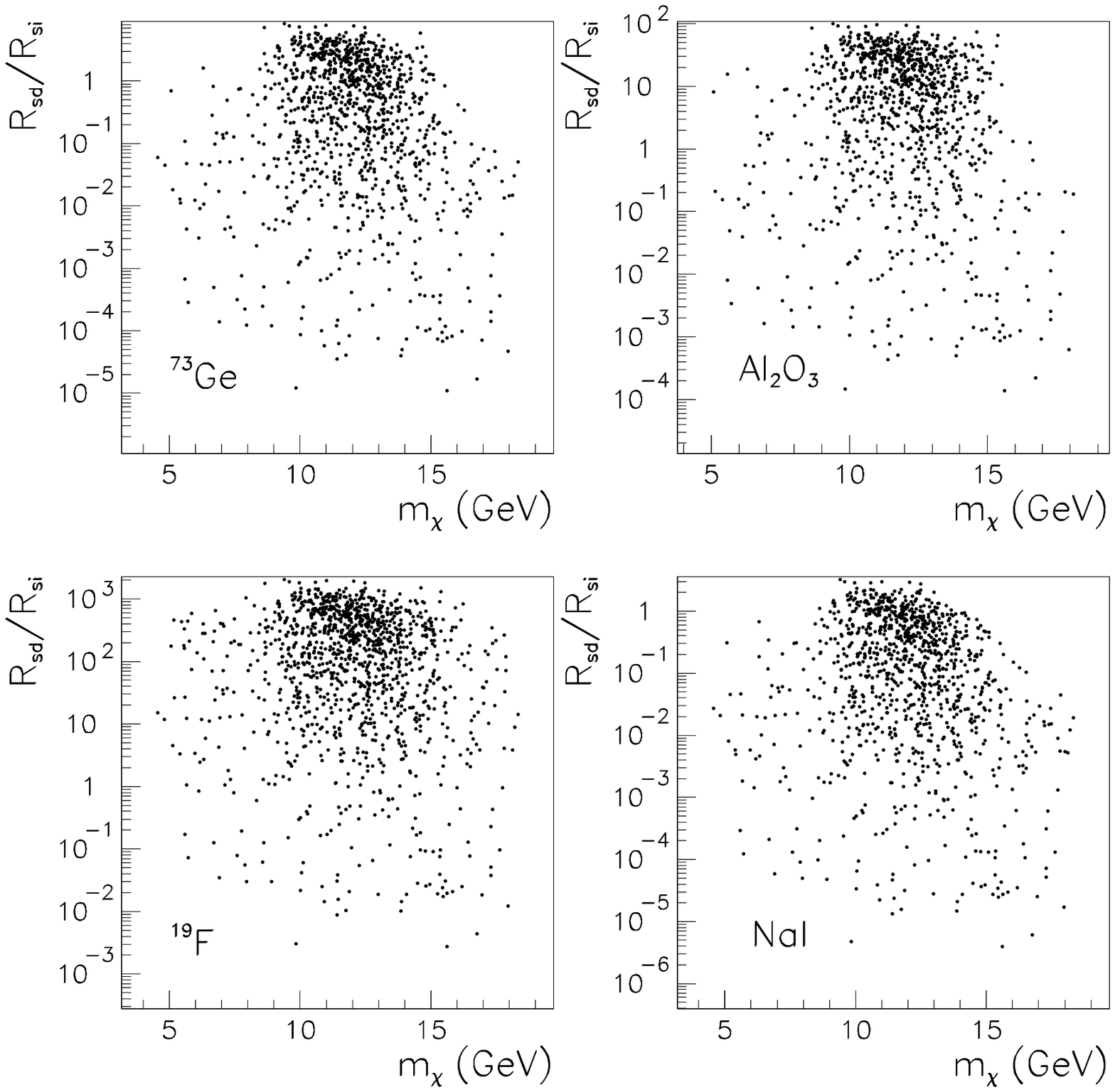}}
\end{picture}

        FIG. 5.
        The ratio $R_{sd}/R_{si}$ versus the neutralino mass $m_{\chi}$.
        $R_{sd}$ and $R_{si}$ are
        the spin-dependent and spin-independent components of $R$
        ($R=R_{sd}+R_{si}$).
        Large event rate GUT scenario (b).

\end{figure} %  voffset=100 -> Up
%%%%%%%%%%%%%%%%%%%%%%%%%%%%%%%%%%%%%%%%%%%%%%%%%%%%%%%%%%%

        An important question touches upon the fraction
        of the spin-dependent part $R_{sd}$ of the total event rate
        $R=R_{sd}+R_{si}$.
        From Eqs. 
\rf{Aq1}--\rf{rel1} it is seen
        that  measurement of $R_{sd}$ and $R_{si}$ would give us
        complementary information about the MSSM parameters.
        In the previously investigated mass region
        $m_{\chi}\geq$~18.4~GeV most of the experimentally interesting isotopes
        are more sensitive to the spin-independent part
        $R_{si}$ due to the coherent enhancement effect
        (see 
~\cite{BKK} and references therein).
        The spin-dependent part $R_{sd}$ originates from
        the neutralino-nucleus interaction via spin-spin coupling.
        It is not a coherent interaction with a  whole nucleus
        since only a few nucleons contribute to the nuclear spin.

        In Figures 4 and 5 we give the scatter plots
        in $R$ -- $R_{sd}/R_{si}$ and  $R_{sd}/R_{si}$ -- $m_{\chi}$
        planes for large event rate scenario (b).
        The following conclusions are valid in scenario (a) as well.
        One can see that in the mass region of the superlight neutralinos
        3~GeV~$\leq m_{\chi}\leq$~20~GeV
        the coherent $R_{si}$ component of the event rate dominates
        for all isotopes in question, except the region of very small
        total event rates $R$.
        As seen from Fig.~5, the spin-dependent component $R_{sd}$
        dominates for the majority of the points in the mass range 
	8 GeV$\leq m_{\chi}\leq$ 14 GeV.
        It reflects the fact
        illustrated in Fig.~1 that 
        in this  mass region the neutralino can acquire the largest higgsino
        admixture $1-Z_g \approx $ 0.35.
        As a result,  the Z-boson contribution to $R_{sd}$
        (via the coefficient ${\cal A}_q$ in Eq. \rf{Aq1})
        is enhanced.

        We conclude this discussion with the following remark.
        Nuclear spin does not play an important role in the 
        direct searches for the DM neutralinos in the mass region 
        3 GeV$\leq m_{\chi}\leq$ 18.4 GeV. Previously the similar 
        conclusion was obtained concerning the mass region 
        18.4 GeV$\leq m_{\chi}$
\cite{BKK}. 
        Nevertheless, only a DM detector with a 
        spin-non-zero target nuclei can provide us with information about
        $R_{sd}$ and the corresponding MSSM parameters in ${\cal A}_q$
        (see Eq. \rf{Aq1}).

%%%%%%%%%%%%%%%%%%%%%
\section{Conclusion}%
%%%%%%%%%%%%%%%%%%%%%

        We have analyzed the MSSM parameter space
        taking into account cosmological
        and accelerator constraints including those from
        the radiative \bsg decay.

        It is well known that the MSSM with the universal gaugino
        mass at the GUT scale disfavors neutralinos lighter than 18.4~GeV
\cite{LRoszk,PDB} if the known experimental data are taken into account.

        A central result of the present paper is the conclusion about
        the existence of a substantial domain of the MSSM parameter space
        corresponding to the superlight neutralinos in the mass range
        3 GeV$\leq m_{\chi}\leq$ 18.4 GeV within the GUT scenarios
        with the non-universal gaugino mass.
        In this domain neutralinos are cosmologically viable and produce an
        event rate which is detectable in the near-future 
        experiments with the low-threshold DM detectors.

\bigskip

        When our paper was being prepared,  we found Ref.
\cite{Gabutti}
        where prospects for the superlight DM neutralino detection are also
        considered but within a more phenomenological approach ignoring
        all GUT conditions. Our analysis shows that we need not
        disregard all GUT conditions in Eqs.
\rf{boundary}--\rf{gauge1}
        to get a window for the superlight neutralinos.
        To this end it is necessary and sufficient to relax
        only gaugino mass unification condition.
        Since we know a generic root of the superlight neutralino we
        can conclude that  in a certain sense 
	the experiments searching for the DM neutralinos
        in the mass range $m_{\chi}\geq 18.4$~GeV
        probe the gaugino mass unification at
        the GUT scale.

\bigskip

{\bf	Acknowledgments}

\smallskip

        We thank W. de Boer, R.Ehret, D.I.Kazakov,
        W.Obersulte-Beckmann and Y.Ramachers for fruitful discussions.

        This work was supported in part by Grant 215 NUCLEON
        from the Ministry of Science and Technological Policy of
        Russian Federation.

\clearpage
%%%%%%%%%%%%%%%%%%%%%%%%%%%%

%%%%%%%%%%%%%%%%%%%%%%%%%%%%%%%%%%%%

\begin{thebibliography}{999}
%%%%%%%%%%%%%%%%%%%%%%%%%%%%
\bibitem{rev}
	For references see the review papers: 
	H.-P.Nilles, Phys. Rep. {\bf 110}, 1 (1984); 
	H.E.Haber and G.L.Kane, Phys. Rep. {\bf 117}, 75  (1985); 
	A.B.Lahanas and D.V.Nanopoulos, Phys. Rep. {\bf 145}, 1 (1987);
   	R.Barbieri, Riv. Nuo. Cim. {\bf 11}, 1 (1988).
\bibitem{roskane}
	G.~L. Kane, C.~Kolda, L.~Roszkowski, and J.~D. Wells,
	Phys.~Rev. D {\bf 49}, 6173 (1994); 
%\bibitem{JuKaGr}
	G.Jungman, M.Kamionkowski and K.Griest
                 "Supersymmetric Dark Matter"
                 SU-4240-605, UCSD/PTH 95-02, IASSNS-HEP-95/14,
                 CU-TP-677, June 1995
                  Phys.Reports, (1996).
\bibitem{LRoszk} L.Roszkowski, Phys.Lett. B {\bf 262}, 59 (1990); 
	B {\bf 278}, 147 (1992).
\bibitem{PDB}Review of~Particle~Properties.
	Phys.~Rev.~D {\bf 50}, 1 (1994).
\bibitem{LEP15}
	L.~Rolandi, H.~Dijksta, D.~Strickland and G.~Wilson,
	representing the ALEPH, DELPHI, L3 and OPAL collaborations,
        Joint seminar on the First Results from LEP1.5, CERN,
         Dec.12th, 1995.
\bibitem{Bernabei} For a recent review see
		 R.Bernabei, R.Nuovo Cimento, {\bf 18}, 1 (1995).
\bibitem{CRESST}
	M. B\"uhler et al., (CRESST Coll.)
	to be published in Nucl. Instr. and Meth. A,
	Proceedings of the 6th Int. Workshop on Low Temperature
	Detectors, Beatenberg-Interlaken, Switzerland,
	28 Aug. - 1 Sept.\ 1995 \\
	W. Seidel et al., Proc.\ 5th Int.\ Workshop on
	Low Temperature Detectors, Berkeley, 29 July -- 3 Aug.\ 1993,
	J. Low Temp. Phys. 93 (1993) 797.
\bibitem{EDELWEISS} D.~Yvon et al., (EDELWEISS Coll.) to be published Proc.
	30th Rencontres de Moriond:
	Clastering in the Universe, Les Arcs, France,	11-18 March 1995;\\
	N.~Coron et al., Nucl. Phys. B (Proc. Suppl.) 35 (1994) 169.
\bibitem{Trofimov} V.N.Trofimov, NIM, A {\bf 370}, 168 (1996).
\bibitem{Heidelberg}
	H.V.Klapdor-Kleingrothaus, Proc. Int. Workshop "Double beta
	decay and related topics", Trento, Italy, April 24 - May 5,
	1995, p.3;
	Y.Ramachers et al. (HEIDELBERG-MOSCOW coll.), Proc. Second
	Workshop of "The Dark Side of Universe", Roma, Nov. 13-14, 1995.
%\bibitem{CalPSINa} D.Reusser et al. Phys.Lett. B {\bf 255}, 143 (1991). 
\bibitem{soft}
	L. Girardello and M.T. Grisaru, Nucl. Phys. {\bf B194}, 419 (1984).
\bibitem{KGrLRosz}
	K.Griest and L. Roszkowski,  Phys. Rev. D {\bf 46}, 3309 (1992). 
\bibitem{BKK} V.A.Bednyakov, H.V.Klapdor-Kleingrothaus and S.G.Kovalenko,
                Phys. Lett.  B {\bf 329}, 5 (1994);
                Phys. Rev.  D {\bf 50}, 7128 (1994). 
\bibitem{WBK} W. de Boer, Prog. in Nucl. and Part. Phys. {\bf 33}, 210 (1994):
        W. de Boer, G. Burkart, R. Ehret, W. Oberschulte-Beckmann,
        V.Bednyakov, D.I.Kazakov  and S.Kovalenko,
        IEKAP-KA/95-07, preprint. hep-ph/9507291;
        IEKAP-KA/96-04, preprint. hep-ph79603350.
\bibitem{loopewbr} 
	R.~Arnowitt and P.~Nath. Phys.~Rev. D {\bf 46}, 3981 (1992).
\bibitem{erz}
	J.~Ellis, G.~Ridolfi, and F.~Zwirner.
	Phys.~Lett. B {\bf 257}, 83 (1991);
	Phys.~Lett.~B {\bf 262}, 477  (1991). 
\bibitem{berz}
	A.~Brignole, J.~Ellis, G.~Ridolfi, and F.~Zwirner.
	Phys.~Lett.~B {\bf 271}, 123 (1991).
\bibitem{LEP}
{The LEP Collaborations: ALEPH, DELPHI, L3 and OPAL and the LEP electroweak
  Working Group;}.
	Phys.~Lett.~B {\bf 276}, 247 (1992);
	 Updates are given in CERN/PPE/93-157, CERN/PPE/94-187 and
  	LEPEWWG/95-01 (see also ALEPH 95-038, DELPHI 95-37, 
	L3 Note 1736 and OPAL TN284.
\bibitem{CDF}{CDF Collab.,~F.~Abe, et al.}
		Phys.~Rev.~Lett.~{\bf 74}, 2626 (1995). 
\bibitem{D0}{D0 Collab.,~S.~Abachi, et al.}
		Phys.~Rev.~Lett.~{\bf 74}, 2632 (1995). 
\bibitem{dfs} G.~Degrassi, S.~Fanchiotti, and A.~Sirlin.
		Nucl.~Phys.~{\bf  B351}, 49 (1991).
\bibitem{EJ95}S.~Eidelmann and F.~Jegerlehner, 1995,
		{PSI Preprint PSI-PR-95-1.}
\bibitem{higgslim} {D.~Buskulic et al., ALEPH Coll.}
		Phys.~Lett.~B {\bf 313}, 312 (1993).
\bibitem{sopczak} {A.~Sopczak}.
		CERN-PPE/94-73, Lisbon Fall School 1993.
\bibitem{cleo94} R.~Ammar et~al. CLEO-Collaboration.
	Phys. Rev. Lett. {\bf  74}, 2885 (1995).
\bibitem{BerBorMasRi}
	S. Bertolini, F. Borzumati, A.Masiero, and G. Ridolfi, 
	Nucl. Phys. {\bf B353}, 591 (1991) 591 {and references therein;} 
	{N. Oshimo}, Nucl. Phys. {\bf  B404}, 20 (1993).
        F.M.Borzumati, M.Drees and M.M.Nojiri,
	Phys.Rev. D {\bf 51}, 341 (1995).
	R. Barbieri and G. Giudice, Phys. Lett. B {\bf 309}, 86 (1993);
        R. Garisto and J.N. Ng, Phys. Lett. B {\bf 315}, 372 (1993).
\bibitem{borz} F. Borzumati, Z. Phys. C {\bf 63}, 291 (1995).
\bibitem{alibsg} A.~Ali and C.~Greub,  Z. Phys. C {\bf 60}, 433 (1993).
\bibitem{burasb} A.J.~Buras et~al, Nucl. Phys. {\bf B424}, 374  (1994).
\bibitem{kolb} E.W. Kolb and M.S. Turner, The early Universe,
		Redwood, Addison-Wesley, (1990).
\bibitem{omega} K. Griest and D. Seckel, Phys. Rev. D {\bf 43}, 3191 (1991); 
		P. Nath and R. Arnowitt, 
		Phys. Rev. Lett. {\bf 70}, 3696 (1993); 
		G. Gelmini and P. Gondolo, 
		Nucl. Phys. {\bf B360}, 145 (1991).
\bibitem{drno} M. Drees and M. M. Nojiri, Phys. Rev. D{\bf 47}, 376 (1993).
\bibitem{relic} G. Steigman, K.A. Olive, D.N. Schramm, M.S. Turner,
	Phys. Lett. B{\bf 176}, 33 (1986);
	J. Ellis, K. Enquist, D.V. Nanopoulos and S. Sarkar,
  	Phys. Lett. B{\bf 167}, 457 (1986);
\bibitem{rosdm} L. Roszkowski, Univ. of Michigan Preprint, UM-TH-93-06;
  		UM-TH-94-02.
\bibitem{relictst}
	J. L. Lopez, D.V. Nanopoulos, and H. Pois,
	Phys. Rev. D{\bf 47}, 2468 (1993);
  	J. L. Lopez, D.V. Nanopoulos, and K. Yuan,
	Phys. Rev. D{\bf 48}, 2766  (1993). 
\bibitem{Gelm} G. B. Gelmini, P. Gondolo and E. Roulet, 
		Nucl. Phys. {\bf  B351}, 623 (1991).
\bibitem{Bot}
	A. Bottino, V. de Alfaro, N. Fornengo, G. Mignola and S.  Scopel,
        Astropart. Phys. J. {\bf 2}, 77 (1994);
        A. Bottino, C.  Favero, N. Fornengo, G. Mignola and S.  Scopel,
        Talk at the Workshop on Double Beta Decay and Related Topics,
        Trento, Italy, 1995.
        V.Berezinsky, A.Bottino, J.Ellis, N.Fornengo,
        G.Mignola and S.Scopel, CERN-TH 95-206, preprint, hep-ph/9508249.
\bibitem{GonBer} P.Gondolo and L.Bergstr\"om,
	UUITP -17/95 OUTP -95 38P, October 1995, hep-ph/9510252
\bibitem{EV} J. Engel and  P. Vogel, Phys. Rev. D{\bf 40}, 3132 (1989);
            J. Engel, S.  Pitel and P. Vogel, 
		Int. J. Mod. Phys.  E {\bf 1}, 1 (1992).
\bibitem{EMC} J. Ashman { et al.} EMC collaboration,
                Nucl. Phys. {\bf  B328}, 1 (1989).
\bibitem{SMC2} G. Mallot, talk presented at SMC meeting on internal
			spin structure of the nucleon, Yale University,
			January 1994.
\bibitem{Dqextract} A. Manohar, R. Jaffe, 
		Nucl. Phys. {\bf B337}, 509 (1990).
\bibitem{ChengGasser} T. P. Cheng, Phys. Rev. D {\bf 38}, 2869 (1988);
                H.-Y. Cheng, Phys. Lett. B {\bf 219}, 347 (1989);
        	J. Gasser, H. Leutwyler and M. E. Sainio,
        	Phys. Lett. B {\bf 253}, 252 (1991).
\bibitem{Hatsuda} T. Hatsuda and T. Kunihiro, 
		Nucl. Phys. {\bf  B387}, 705 (1992).
\bibitem{Engel}  J. Engel, Phys. Lett. B{\bf 264}, 114 (1991).
\bibitem{Ressell} M.T. Ressell, M.B. Auferheide, S.D. Bloom,
        K. Griest, G.J. Mathews and D.A. Resler, 
		Phys. Rev. D{\bf 48}, 5519 (1993).
\bibitem{Nikolaev} M. A. Nikolaev, H. V. Klapdor-Kleingrothaus,
                  Z. Phys. A {\bf 345}, 183 (1993); ibid, 373 (1993). 
\bibitem{EF}  J. Ellis, R. Flores, Phys. Lett. B{\bf 263}, 259 (1991);
                Phys. Lett.  B {\bf 300}, 175 (1993); 
		Nucl. Phys.  {\bf B400}, 2536 (1993).
\bibitem{Gabutti}
	A.Gabutti, M.Olechovski, S.Cooper, S.Pokorcki and L.Stodolsky
      "Light Neutralinos as Dark Matter in the Unconstrained MSSM "
 	MPI-PHE/96-02; DFTT 1/96 (hep-ph/9602432)

\end{thebibliography}
\end{document}